\documentclass[pre,aps,amsmath,amssymb,amsfonts,floatfix,superscriptaddress,showpacs,twocolumn,footinbib]{revtex4-1}
\usepackage{graphicx}
\usepackage{amsmath,amsfonts}
\usepackage{xcolor}
\usepackage{color}
\usepackage{mathtools}
\usepackage{eufrak}
\usepackage{pgfplots}
\usepackage{soul}
\usepackage{multirow}
\usepackage{comment}
\usepackage{ulem}

\newcommand{\up}{\uparrow}
\newcommand{\dn}{\downarrow}
\newcommand{\kv}{\ensuremath{\mathbf{k}}}
\newcommand{\qv}{\ensuremath{\mathbf{q}}}

\newcommand{\SBE}{\ensuremath{\text{SBE}}}
\newcommand{\ch}{\ensuremath{\text{ch}}}
\newcommand{\sz}{\ensuremath{\text{sp}}}
\newcommand{\av}[1]{\ensuremath{\left\langle #1 \right\rangle}}

\newcommand{\sing}{\ensuremath{\text{s}}}

\newcommand{\Uirr}{\ensuremath{\text{Uirr}}}

\usepackage{tikz}

\usetikzlibrary{calc}
\usetikzlibrary{decorations.pathmorphing}
\usetikzlibrary{decorations.pathreplacing}
\usetikzlibrary{decorations.markings}
\usetikzlibrary{shapes.misc}
\usetikzlibrary{shapes}
\usetikzlibrary{positioning}
\usetikzlibrary{snakes}
\usetikzlibrary{arrows}
\usetikzlibrary{fadings}
\usetikzlibrary{calc,arrows}
\tikzstyle{decision} = [diamond, draw, fill=blue!20, text width=4.5em, text badly centered, node distance=3cm, inner sep=0pt]
\tikzstyle{block} = [rectangle, draw, fill=blue!20, text width=7em, text centered, rounded corners, minimum height=5em]
\tikzstyle{line} = [draw, -latex']
\tikzstyle{cloud} = [draw, ellipse,fill=red!20, node distance=3cm, minimum height=2em]
\tikzstyle{overbrace style}=[decorate,decoration={brace,raise=2mm,amplitude=3pt}]
\tikzstyle{overbrace text style}=[font=\footnotesize, above, pos=.5, yshift=3mm]
\tikzset{snake it/.style={decorate, decoration=snake}}
\usetikzlibrary{3d}
\usepackage{mathtools}
    \tikzset{
            partial ellipse/.style args={#1:#2:#3}{
                        insert path={+ (#1:#3) arc (#1:#2:#3)}
                            }
                        }

\usetikzlibrary{calc}
\tikzset{
            inertial frame/.style = {x={(-20:2cm)}, y={(-160:2cm)}, z={(90:2cm)}},
              local frame/.style = {shift={(local origin)}, x={(40:.7cm)}, y={(150:.7cm)}, z={(105:.7cm)}}
          }

    \tikzset{middlearrow/.style={
                decoration={markings,
                            mark= at position 0.65 with {\arrow{#1}} ,
                                    },
                                            postaction={decorate}
                                                }
                                                }
\tikzset{cross/.style={cross out, draw, 
         minimum size=2*(#1-\pgflinewidth), 
                  inner sep=0pt, outer sep=0pt}}

\def\presuper#1#2%
  {\mathop{}%
   \mathopen{\vphantom{#2}}^{#1}%
   \kern-0.5\scriptspace%
   #2}

\usepackage{hyperref}
\hypersetup{colorlinks=true,breaklinks,linkcolor=blue,urlcolor=blue,citecolor=blue}
\begin{document}

    \pgfmathdeclarefunction{gauss}{2}{%
          \pgfmathparse{1/(#2*sqrt(2*pi))*exp(-((x-#1)^2)/(2*#2^2))}%
          }
    \pgfmathdeclarefunction{mgauss}{2}{%
          \pgfmathparse{-1/(#2*sqrt(2*pi))*exp(-((x-#1)^2)/(2*#2^2))}%
          }
    \pgfmathdeclarefunction{lorentzian}{2}{%
        \pgfmathparse{1/(#2*pi)*((#2)^2)/((x-#1)^2+(#2)^2)}%
          }
    \pgfmathdeclarefunction{mlorentzian}{2}{%
        \pgfmathparse{-1/(#2*pi)*((#2)^2)/((x-#1)^2+(#2)^2)}%
          }

\author{Viktor Harkov}
\affiliation{Institute of Theoretical Physics, University of Hamburg, 20355 Hamburg, Germany}
\affiliation{European X-Ray Free-Electron Laser Facility, Holzkoppel 4, 22869 Schenefeld, Germany}
\author{Alexander I. Lichtenstein}
\affiliation{Institute of Theoretical Physics, University of Hamburg, 20355 Hamburg, Germany}
\affiliation{European X-Ray Free-Electron Laser Facility, Holzkoppel 4, 22869 Schenefeld, Germany}
\author{Friedrich Krien}
\affiliation{Institute for Solid State Physics, TU Wien, 1040 Vienna, Austria}
\affiliation{Jo\v{z}ef Stefan Institute, Jamova 39, SI-1000, Ljubljana, Slovenia}

\title{Parametrizations of local vertex corrections from weak to strong coupling:\\
importance of the Hedin three-leg vertex}

\begin{abstract}
In the study of correlated systems, approximations based on the dynamical mean-field theory (DMFT) provide a practical way to take local vertex corrections into account, which capture, respectively, particle-particle screening at weak coupling and the formation of the local moment at strong coupling. We show that in both limits the local vertex corrections can be efficiently parametrized in terms of single-boson exchange, such that the two-particle physics described by DMFT and its diagrammatic extensions is recovered to good approximation and at a reduced computational cost. Our investigation highlights the importance of the frequency-dependent fermion-boson coupling (Hedin vertex) for local vertex corrections. Namely, at weak coupling the fermion-spin-boson coupling suppresses the N\'eel temperature of the DMFT approximation compared to the static mean-field, whereas for large interaction it facilitates a huge enhancement of local spin-fluctuation exchange, giving rise to the effective-exchange energy scale $4t^2/U$. We find that parametrizations of the vertex which neglect the nontrivial part of the fermion-boson coupling fail qualitatively at strong coupling.
\end{abstract}

\maketitle

Two-particle electronic correlations provide an essential and complementary viewpoint on correlated systems~\cite{Rohringer12}. For example, the calculation of susceptibilities greatly simplifies the study of second order phase transitions, compared to calculations on the one-particle level which require the introduction of complicated unit cells and conjugate fields~\cite{Boehnke12,Geffroy19,Strand19,Pavarini21}. Further, in a great variety of perturbative many-body techniques the electronic self-energy is given as a set of vertex diagrams, so that in effect two-particle quantities need to be evaluated before one obtains information about the one-particle correlations~\cite{Rohringer18}. In fact, even the celebrated Fermi liquid theory is at its core defined through two-particle scattering amplitudes, the Landau parameters~\cite{Landau56,Noziere97,Krien19-2,Melnick20}.

However, two problems are often encountered in the study of two-particle correlations. First, even in case of simple systems, it is exceedingly difficult to compute or even only memorize~\cite{Kauch17} on present-day computing devices the complete two-particle information, as represented by the four-point vertex function. Second, a clear organizing principle for two-particle correlations is often missing, which would allow to decode the physics encapsulated in the vertex function.

Here we like to put forth the notion, and demonstrate in practice on a particular example, that the two mentioned problems are in fact related: When the two-particle correlations are organized in a way that mirrors the key physics at play, the computational effort can be reduced. Or one can more easily define suitable approximations which achieve this goal. 

We derive this guiding theme from both established and recently introduced (partial) bosonizations of the vertex function~\cite{Krahl07,Friederich10,Denz20}. Namely, in the context of the functional renormalization group (fRG,~\cite{Metzner12,Dupuis21}) the vertex is often decomposed into various channels~\cite{Karrasch08,Wang12,Vilardi17,Vilardi19,Tagliavini19,Wentzell20,Kugler18,Kugler18-2}, or parametrized in terms of bosonic fluctuations and their (Yukawa) coupling to fermions~\cite{Husemann09,Husemann12,Eberlein_thesis,Bonetti20}. In particular the latter procedure provides, simultaneously, and in accord with the guiding theme of this work, (i) an organizing principle for two-particle correlations and (ii) a reduction of the computational effort. This is because, on the one hand, the bosonic propagators and Yukawa couplings have a transparent physical interpretation, and, on the other hand, they can be computed and stored more easily than a genuine four-point vertex.

In this context, it can be shown under very general assumptions~\footnote{For a description of these general assumptions see Appendix E of Ref.~\cite{Krien19}.} that vertex diagrams have special properties when they are reducible with respect to the interaction. Namely, interaction-reducible diagrams give rise to the high-frequency asymptotics of the vertex function~\cite{Wentzell20}, and in the case of a lattice model, they carry crucial information about its dependence on the momenta~\cite{Krien20,Krien20-2,Krien20-3}. In the single-boson exchange (SBE) decomposition~\cite{Krien19-4} the vertex diagrams are grouped into three interaction-reducible classes and one irreducible class. Each reducible class corresponds to the exchange of a single boson and the irreducible class represents multiple boson exchange~\cite{Krien20,Krien20-2}. In this formalism the bosons and Yukawa couplings can be identified, respectively, as the screened interaction and the Hedin three-leg vertex of the $GW\gamma$ theory~\cite{Hedin65}. In this way, the SBE decomposition provides a conceptual link between Hedin's equations~\cite{Krien20-2}, vertex asymptotics~\cite{Wentzell20}, partial bosonizations~\cite{Denz20,Gunnarsson15,Stepanov19-2,Harkov21}, and the parquet approach~\cite{Diatlov57,Dominicis64-2,Bickers04}.

Here we apply this general framework to a specific problem, namely, the calculation of two-particle correlation functions within dynamical mean-field theory~\cite{Georges96} and its diagrammatic extensions~\cite{Rohringer18}. These methods rely on the solution of an auxiliary Anderson impurity model (AIM) which provides the local correlations non-perturbatively. Here, the computationally most expensive step is in general the evaluation of the local vertex function $f(\nu,\nu',\omega)$ of the impurity model, for example, using continuous-time quantum Monte Carlo (CTQMC) algorithms~\cite{Gull11}, due to its dependence on one bosonic ($\omega$) and two fermionic ($\nu,\nu'$) Matsubara frequencies. This step is especially expensive in multi-orbital settings, where the calculation of a single vertex function may require hundreds of thousands of core hours~\cite{Galler19}, raising questions regarding the feasibility of frequent calculations and parameter scans.

\begin{figure}
     \begin{center}
 	\begin{tikzpicture}
    \node[anchor=south west,inner sep=0] (image1) at (0,0)
    {\includegraphics[width=0.45\textwidth]{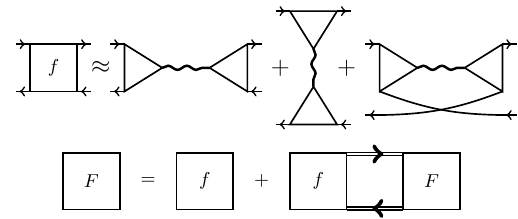}};
	\end{tikzpicture}
    \end{center}
    \vspace{-.5cm}
     \caption{\label{fig:jib} Top: Symbolic representation of the {\sl SBE-approximation}.
     Wiggly lines and triangles represent, respectively, the screened interaction $w$ and the Hedin vertices $\lambda$. The {\sl $w$-approximation} corresponds to setting $|\lambda|=1$.
     Arrows denote the impurity Green's function $g$;
     prefactors, flavor labels, and a double counting correction are omitted.
     Bottom: The nonlocal Bethe-Salpeter equation constructs the lattice vertex function $F$
     from impurity vertex $f$ and nonlocal propagator $\tilde{G}$ (double arrows).
     }
     \end{figure}

We address this problem in the spirit of the discussed strategy to first find a useful organizing principle for the two-particle correlations, and second to define a cheap approximation for the impurity vertex function based on this insight. To this end, we employ the approximation for the impurity vertex sketched on the top of Fig.~\ref{fig:jib}, where the vertex is parametrized in terms of the screened interaction (wiggly lines) and the Hedin vertex (triangles), which are computationally cheap to obtain. On the other hand, the approximation neglects a residual (interaction-irreducible) four-point vertex `$\varphi^\text{Uirr}$', whose calculation is computationally expensive. 

In the following, we show that this `{SBE approximation}' is {\sl sufficient} in the sense that it recovers the two-particle physics described by DMFT and its extensions at both weak and strong coupling. But we also show that it is {\sl necessary} to keep this much information about the impurity vertex function. In particular, the nontrivial (interacting) part of the Hedin vertex should not be neglected, because it leads to qualitatively wrong results at strong coupling. In this limit parametrizations based only on the screened interaction or on vertex asymptotics fail. In the latter case, we show that this can be related to qualitative differences in the frequency structure of the residual vertices (also called rest functions~\cite{Wentzell20}) of, respectively, the SBE decomposition and the vertex asymptotics.

For concreteness, we consider the paramagnetic Hubbard model on the square lattice at half-filling,
\begin{align}
    H = &-\sum_{\langle ij\rangle\sigma}{t}_{ij} c^\dagger_{i\sigma}c^{}_{j\sigma}+ U\sum_{i} n_{i\up} n_{i\dn},\label{eq:hubbard}
\end{align}
where $t_{ij}$ denotes the hopping between nearest neighbors i and j,
its absolute value $t=1$ sets the unit of energy.
$c^{},c^\dagger$ are the annihilation and creation operators with the spin index $\sigma=\up,\dn$.
The Coulomb repulsion between the densities $n_{\sigma}=c^\dagger_{\sigma}c^{}_{\sigma}$ is denoted by $U$. 

\section*{Local vertex corrections}
In DMFT~\cite{Georges96} the Hubbard model~\eqref{eq:hubbard} is mapped to the auxiliary AIM with the action,
\begin{align}
  S_{\text{AIM}}=&-\sum_{\nu\sigma}c^*_{\nu\sigma}(\imath\nu+\mu-\Delta_\nu)c^{}_{\nu\sigma}+U \sum_\omega n_{\up\omega} n_{\dn\omega},
  \label{eq:aim}
\end{align}
where $c$ and $c^*$ are Grassmann numbers and $\nu$ and $\omega$ are fermionic and bosonic Matsubara frequencies, respectively. Half-filling is implied by setting the chemical potential to $\mu=\frac{U}{2}$ to enforce particle-hole symmetry. Summations over Matsubara frequencies $\nu, \omega$ contain implicitly the factor $T$, the temperature. The DMFT hybridization function $\Delta_\nu$ is fixed by the self-consistency condition, $G_{ii}(\nu)=g(\nu)$, for the local Green's function $G_{ii}$ of the Hubbard model (\ref{eq:hubbard}) and the Green's function of the impurity model (\ref{eq:aim}), $g_{\sigma}(\nu)=-\langle c_{\nu\sigma}c^{*}_{\nu\sigma}\rangle$. The label $\sigma$ is suppressed where unambiguous. 

In the DMFT approximation the {\sl nonlocal} two-particle correlations arise from a Bethe-Salpeter equation with {\sl local} vertex corrections~\cite{Georges96,Toschi07}. In the equivalent dual fermion formulation~\cite{Rubtsov08,Hafermann14-2,vanLoon20} one avoids the (particle-hole) irreducible vertex and instead builds the vertex corrections from the full vertex function $f$ of the AIM~\eqref{eq:aim}, cf. Fig.~\ref{fig:jib} bottom,
\begin{align}
  F^{\alpha}_{\nu\nu'}(q)=f^{\alpha}_{\nu\nu'\omega}+\sum_{\nu''}f^{\alpha}_{\nu\nu''\omega}\tilde{X}^{0}_{\nu''}(q)F^{\alpha}_{\nu''\nu'}(q).
  \label{eq:BSE}
\end{align}
Here, $\tilde{X}^{0}_{\nu}(q)=\sum_{\textbf{k}}\tilde{G}_{k}\tilde{G}_{k+q}$ is a bubble of nonlocal propagators $\tilde{G}$, where $k=(\kv,\nu), q=(\qv,\omega)$, and the summation over the momentum $\kv$ implies division by the number of lattice sites $N$. For the precise definition of the impurity vertex function $f$ see Appendix~\ref{app:defs}. At DMFT level $\tilde{G}={\cal G}$, where ${\cal G}_k\equiv G_{k}-g_{\nu}$ is the nonlocal DMFT Green's function, whereas in the dual fermion approach $\tilde{G}$ is further dressed with a self-energy~\cite{Rubtsov08}, $\tilde{G}_k^{-1}={\cal G}^{-1}_k-\tilde{\Sigma}_k$. In the following, Eq.~\eqref{eq:BSE} serves us as a generic starting point to evaluate the two-particle correlations within DMFT ($\tilde{\Sigma}=0$) and in the ladder dual fermion approach (LDFA,~\cite{Hafermann09}; with $\tilde{\Sigma}$ given as described further below).

In our applications we evaluate Eq.~\eqref{eq:BSE} with a physical cutoff $\nu^\text{max}=45t$ for the fermionic Matsubara frequencies $\nu_n$, that is, a variable grid with $-\left \lfloor{\nu^\text{max}/(2\pi T)}\right \rfloor-1\leq n\leq\left \lfloor{\nu^\text{max}/(2\pi T)}\right \rfloor$, and similarly $0\leq m\leq\left \lfloor{\nu^\text{max}/(\pi T)}\right \rfloor$ for bosonic frequencies $\omega_m$. The lattice size is $64\times 64$.

\section*{Vertex parametrizations}
The Bethe-Salpeter Eq.~\eqref{eq:BSE} above requires the local vertex function $f$ of the AIM~\eqref{eq:aim} as an input. In this work we focus on the two-particle level of DMFT and on the LDFA, however, this requirement is quite general. Indeed, all diagrammatic extensions of DMFT~\cite{Rohringer18} which take local four-point vertex corrections into account require, by construction, knowledge of $f$. However, this is connected with the calculation of the four-point correlation function of the AIM,
\begin{align*}
f^\alpha(\nu,\nu',\omega)\propto\sum_{\sigma_i}s^{\alpha}_{\sigma'_{1}\sigma^{}_{1}}s^{\alpha}_{\sigma'_{2}\sigma^{}_{2}}\langle c^{}_{\nu\sigma^{}_{1}}c^{*}_{\nu+\omega,\sigma'_{1}}c^{}_{\nu'+\omega,\sigma^{}_{2}}c^{*}_{\nu'\sigma'_{2}}\rangle,
\end{align*}
where $s^\alpha$ are the Pauli matrices ($\alpha=\ch$ or $\alpha=x,y,z\equiv\sz$) and $\langle ...\rangle$ denotes an impurity average, see also Appendix~\ref{app:defs}.

Previously, this correlation function has been calculated using a variety of different methods, for example, full exact diagonalization (ED,~\cite{Toschi07}), Lanczos ED~\cite{Tanaka19}, CTQMC, and, very recently, the numerical renormalization group (NRG,~\cite{Lee21,Kugler21}). All these methods have in common that the evaluation of $f$ is computationally expensive, in fact, at the present stage only CTQMC solvers are applicable in multi-orbital settings~\cite{ALPS2,Wallerberger19}. To alleviate this problem we investigate in the following approximations for the impurity vertex function $f$, which do not require the calculation of the four-point correlation function of the AIM. A suitable approximation should recover the two-particle physics described by DMFT and its extensions, if not quantitatively, at least qualitatively.

\begin{figure*}
\includegraphics[scale=0.9]{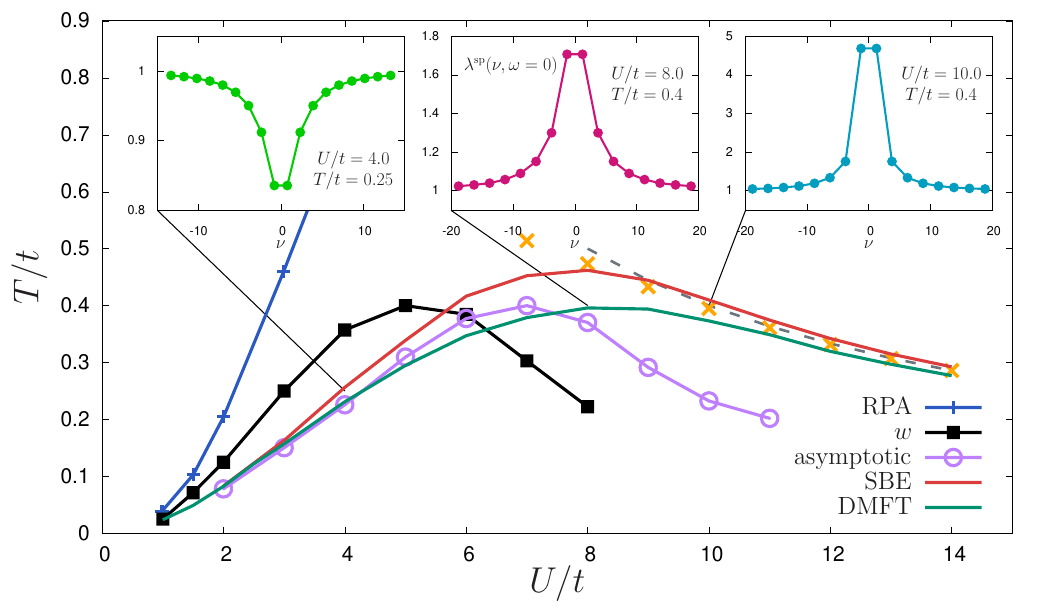}
\caption{{Main panel:} {N\'eel temperature of the half-filled Hubbard model~\eqref{eq:hubbard}} in various approximations (see text). A dashed line indicates $4t^2/U=T$; orange crosses show points where $J=T$, with the effective exchange $J$ calculated from Eq.~\eqref{eq:jeff}. {Insets:} Local fermion-spin-boson coupling of the self-consistent AIM corresponding to indicated $(U,T)$ in main panel.}
\label{fig:phase_diag}
\end{figure*}

To this end, we make use of the fact that the vertex $f$ can be decomposed into single-boson exchange (SBE) diagrams and a residual four-fermion vertex~\cite{Krien19-4},
\begin{align}
    f^\alpha_{\nu\nu'\omega}=\varphi^{\Uirr,\alpha}_{\nu\nu'\omega}+\nabla^{\SBE,\alpha}_{\nu\nu'\omega}.
    \label{eq:ff-vertex}
\end{align}
Only the residual vertex $\varphi^{\Uirr}$ is intrinsically a four-point quantity, whereas the single-boson exchange can be constructed from the screened interaction $w^\alpha(\omega)$ and the Hedin vertex $\lambda^\alpha(\nu,\omega)$ of the AIM, where $\alpha=\ch,\sz,\sing$ denotes the charge, spin, or singlet flavor.

The measurement of these quantities requires much less computational resources than the measurement of the four-point vertex because they can be expressed, respectively, in terms of two- and three-point correlation functions of the AIM (cf. Appendix~\ref{app:defs},~\cite{vanLoon18}),
\begin{align*}
w^{\alpha}_\omega\propto&\av{\rho^{\alpha}_{-\omega}\rho^{\alpha}_{\omega}},\;\;\; \lambda^{\alpha}_{\nu,\omega}\propto\sum_{\sigma\sigma'}s^{\alpha}_{\sigma'\sigma}\av{c_{\nu\sigma}c^{*}_{\nu+\omega\sigma'}\rho^{\alpha}_{\omega}},\\
w^{\sing}_\omega\propto&\av{\rho^{+}_{-{\omega}}\rho^{-}_{{\omega}}},\;\;\;
{\lambda}^{\sing}_{\nu,\omega}\propto\av{c_{\nu\uparrow}c_{{\omega}-\nu,\downarrow}\rho^{+}_{{\omega}}},
\end{align*}
where in the first line $\alpha=\ch,\sz$ and $\rho^{\ch}=n_{\uparrow}+n_{\downarrow}$ and $\rho^{\sz}=n_{\uparrow}-n_{\downarrow}$ are the charge and spin densities; in the second line $\rho^{+}=c^{*}_{\uparrow}c^{*}_{\downarrow}$ and $\rho^{-}=c_{\downarrow}c_{\uparrow}$ describe creation and annihilation of electron pairs, respectively. 

In terms of $w$ and $\lambda$ the SBE diagrams $\nabla^{\SBE}$ in Eq.~\eqref{eq:ff-vertex} are given as the sum of three SBE contributions and a double counting correction $2U$, for $\alpha=\ch,\sz$ it reads,
\begin{align}
    \nabla^{\SBE,\alpha}_{\nu\nu'\omega}=\nabla^{ph,\alpha}_{\nu\nu'\omega}
    +\nabla^{\overline{ph},\alpha}_{\nu\nu'\omega}+\nabla^{pp,\alpha}_{\nu\nu',\omega+\nu+\nu'}-2U^\alpha,
    \label{eq:SBEvertex}
\end{align}
where $U^\ch=+U, U^\sz=-U$ is the bare interaction in the respective channel and the $\nabla$'s are defined as,
\begin{subequations}
 \label{eq:SBEvertices_all}
\begin{align}
    \nabla^{ph,\alpha}_{\nu\nu'\omega}&\!=\!{\lambda}^{\alpha}_{\nu\omega} w_{\omega}^{\alpha}\lambda_{\nu'\omega}^{\alpha},\label{eq:SBEvertices_ph}\\
    \nabla^{\overline{ph},\alpha}_{\nu\nu'\omega}&\!=\!
    -\frac{1}{2}\nabla^{ph,\ch}_{\nu,\nu+\omega,\nu'-\nu}
    \!-\!\frac{3\!-\!4\delta_{\alpha,\sz}}{2}\nabla^{ph,\sz}_{\nu,\nu+\omega,\nu'-\nu},\label{eq:SBEvertices_vph}\\
    \nabla^{pp,\alpha}_{\nu\nu'\omega}&\!=\!\frac{1-2\delta_{\alpha,\sz}}{2}{\lambda}^{\sing}_{\nu{\omega}}w_{{\omega}}^{\sing}\lambda_{\nu'{\omega}}^{\sing}.
    \label{eq:SBEvertices}
\end{align}
\end{subequations}

In the following we investigate parametrizations of the impurity vertex function based on SBE diagrams, where we test the following three approximations:
\begin{itemize}
\item In the `{\sl SBE approximation}'~\cite{Krien19-4} we keep the diagrams shown symbolically on the top of Fig.~\ref{fig:jib},
\begin{align}
    f^{\alpha}_{\nu\nu'\omega}\approx\nabla^{\SBE,\alpha}_{\nu\nu'\omega}.\label{eq:sbe_approximation}
\end{align}
The organizing principle here is to parametrize the vertex in terms of bosonic fluctuations $w(\omega)$ and their coupling $\lambda(\nu,\omega)$ to the fermions.
\item The `{\sl $w$ approximation}' is similar to the SBE approximation but we simplify also the fermion-boson couplings, setting $\lambda^{\ch/\sz}=1$ and $\lambda^\sing=-1$. As a result, this approximation requires only knowledge of the screened interaction $w(\omega)$. Here the organizing principle is to parametrize the vertex in terms of bosonic fluctuations only. Similar approximations were introduced in Refs.~\cite{Kusunose10,Kunes11,Tagliavini18}.
\item In the `{\sl asymptotic approximation}' the vertex $f$ is parametrized based on its asymptotic value when one or multiple frequencies~\footnote{One needs to consider separately the limits where only $|\nu|\rightarrow\infty$ or $|\nu'|\rightarrow\infty$; or $|\nu|$ and $|\nu'|\rightarrow\infty$ but either $\nu'-\nu=\text{const}$ or $\nu'+\nu+\omega=\text{const}$, see also Ref.~\cite{Wentzell20}.} are large~\cite{Wentzell20}. To be unbiased with respect to the limit that is taken, the asymptotic expressions are combined so that the correct vertex asymptote is recovered in each limit~\cite{Tagliavini18}. The asymptotic approximation can be obtained from the SBE approximaton by replacing
\begin{align}
\nabla^{\alpha}_{\nu\nu'\omega}=&{\lambda}^{\alpha}_{\nu\omega} w_{\omega}^{\alpha}\lambda_{\nu'\omega}^{\alpha}\notag\\
\approx&{\lambda}^{\alpha}_{\nu\omega} w_{\omega}^{\alpha}(\pm1)
    +(\pm1)w_{\omega}^{\alpha}\lambda_{\nu'\omega}^{\alpha}
    -w_{\omega}^{\alpha},\label{eq:asym}
\end{align}
where $+$ corresponds to $\alpha=\ch,\sz$ and $-$ to $\alpha=\sing$, and inserting into Eq.~\eqref{eq:sbe_approximation} using Eqs.~\eqref{eq:SBEvertex} and~\eqref{eq:SBEvertices_all}. The organizing principle here is, of course, to recover the vertex asymptotically. Since the asymptotic approximation can be derived from the SBE approximation they become equivalent at high frequencies. This parametrization is used frequently in fRG schemes, combined with a more elaborate treatment of the corresponding rest function~\cite{Wentzell20}.
\end{itemize}

The three approximations defined above all have in common that they do not require the four-point vertex $f$ of the AIM. The $w$ approximation requires even only the screened interaction as an input and is therefore computationally very cheap. It is motivated by the fact that the asymptotic value of the Hedin vertex is unity,
\begin{align}
\lim\limits_{|\nu|\rightarrow\infty}\lambda^{\ch/\sz}_{\nu\omega}=1,\;\;\;
\lim\limits_{|\nu|\rightarrow\infty}\lambda^{\sing}_{\nu\omega}=-1.
\end{align}
Therefore, the asymptotic approximation lies in between the SBE approximation and the $w$ approximation, because of the two fermion-boson couplings $\lambda$ one is kept and the other is set to its asymptotic value, cf. Eq.~\eqref{eq:asym}.

\section*{DMFT -- Phase diagram}
To benchmark the three approximations defined in the previous section, we first calculate the antiferromagnetic phase boundary of the Hubbard model~\eqref{eq:hubbard}.

In fact, in two dimensions this phase transition is forbidden by the Mermin-Wagner theorem. Nevertheless, the phase boundary predicted by DMFT is a useful indicator for the region of the phase diagram with strong spin fluctuations~\cite{Rohringer18}. Furthermore, it has been shown in various studies that even below the critical temperature of DMFT both the paramagnetic and the ordered DMFT solution capture remarkably well many aspects of the phases that can be realized in accord with the Mermin-Wagner theorem or in the experiment. This includes local observables~\cite{Schaefer21}, the effective exchange $J$ at strong coupling~\cite{Kiani16,Stepanov18,Krien19}, commensurate and incommensurate spin-density waves out of half-filling~\cite{Vilardi18}, Fermi surface reconstruction~\cite{Bonetti20-2}, and even stripe order~\cite{Peters14}. It is therefore of a vital interest to gain insight into the mechanism of the N\'eel order predicted by DMFT and to identify its key ingredients~\cite{delRe21}.

Fig.~\ref{fig:phase_diag} shows the various phase boundaries obtained from the random phase approximation (RPA, blue line) and DMFT, where the green line is obtained from the leading eigenvalue of the Bethe-Salpeter kernel [cf. Eq.~\eqref{eq:BSE}],
\begin{align}
\sum_{\nu''}f^{\sz}_{\nu\nu''\omega=0}\tilde{X}^{0}_{\nu''}(\mathbf{q}=(\pi,\pi),\omega=0),\label{eq:matrix_ev}
\end{align}
where $\tilde{X}^0$ is the nonlocal bubble defined below Eq.~\eqref{eq:BSE} (with $\tilde{\Sigma}=0$). At the phase transition the leading eigenvalue approaches unity. Note that here $f$ still corresponds to the numerically exact four-point vertex of the self-consistent AIM~\eqref{eq:aim}.

The comparison of DMFT and RPA at weak coupling highlights the fact that, in two dimensions, the exponential N\'eel temperature $\propto\!\exp(-2\pi\sqrt{t/U})$ of the RPA is {\sl nowhere} quantitatively predictive of the onset of strong spin fluctuations. This is instead the case for DMFT which includes an important vertex correction due to particle-particle (Kanamori) screening, leading to renormalization of the exponent~\cite{Georges91,Kopietz93,Heiselberg00,Toschi05}. It has been emphasized that this vertex correction is included in the fermion-spin-boson coupling~\cite{Katanin09,Krien20,Krien20-2}. For large $U\gtrsim12$ the N\'eel temperature of DMFT inflects towards its strong-coupling asymptote $4t^2/U$ (dashed line)~\cite{Georges96}.

Next, we apply the parametrizations of the local vertex corrections defined in the previous section and estimate the leading eigenvalue of the Bethe-Salpeter kernel~\eqref{eq:matrix_ev} using the corresponding approximation for the impurity vertex $f$. First, using the $w$ approximation we obtain the black curve in Fig.~\ref{fig:phase_diag}. Apparently, this approximation deviates from the original DMFT curve already at weak coupling and fails qualitatively at strong coupling, in fact, its N\'eel temperature drops sharply already at intermediate $U$. Second, we consider the asymptotic approximation represented by the purple curve in Fig.~\ref{fig:phase_diag}. This approximation agrees well with the original DMFT phase boundary for weak coupling. However, at intermediate coupling the N\'eel temperature drops similar to the $w$ approximation and shows an inflection point near $U/t\approx9$ that connects to a strong-coupling asymptote $\propto1/U$ with a coefficient different from $4t^2$. Finally, the red curve shows the SBE approximation, which is quantitatively accurate both in the weak and strong coupling limits and agrees qualitatively with the original DMFT curve throughout the phase diagram. In terms of quantitative accuracy the asymptotic approximation lies closer to DMFT for weak to intermediate couplings, which seems to be a result of its overall smaller N\'eel temperature compared to the SBE approximation.

As shown in Ref.~\cite{Krien19-4}, the SBE approximation can be justified in the weak coupling limit~\footnote{Other than for the square lattice Hubbard model~\eqref{eq:hubbard} itself, a weak-coupling expansion can be applied to the AIM~\eqref{eq:aim} corresponding to the DMFT solution of this model at small $U$~\cite{Rohringer12}.} because it recovers all diagrams for $f$ up to order $\mathcal{O}(U^3)$. However, the agreement with the full DMFT solution at strong coupling requires a further explanation. Indeed, it has been shown for the ordered~\cite{Stepanov18} and for the paramagnetic~\cite{Krien19} DMFT solution that at strong coupling it is sufficient to approximate $f^\sz\approx\nabla^{ph,\sz}$, where $\nabla^{ph,\sz}$ is defined in Eq.~\eqref{eq:SBEvertices_ph}. On the top of Fig.~\ref{fig:jib} this corresponds to keeping only the first diagram on the right-hand-side. Using this approximation the effective exchange is given as~\cite{Krien19}
\begin{align}
J=\frac{2t^2}{(\pi^\sz_{\omega=0})^2}\sum_\nu \lambda^\sz_{\nu,\omega=0}(g_\nu)^4\lambda^\sz_{\nu,\omega=0},\label{eq:jeff}
\end{align}
where $\pi^\sz$ is the spin polarization of the impurity. As shown in Ref.~\cite{Krien19}, for large $U$ this expression approaches ${4t^2}/{U}$ (see orange crosses in Fig.~\ref{fig:phase_diag}). Since $\nabla^{ph}$ corresponds to a subset of the SBE diagrams, the SBE approximation also recovers the effective exchange in this limit. Importantly, the Hedin vertex $\lambda^\sz$ makes a large contribution to Eq.~\eqref{eq:jeff} and hence approximations may fail to recover the energy scale $J$ when $\lambda^\sz$ is set to $1$ in some place, as is confirmed by the results shown in Fig.~\ref{fig:phase_diag}.

Notice that, while the SBE approximation recovers the strong-coupling limit exactly, this does not imply that the residual vertex $\varphi^\text{\Uirr,\sz}$ is small. Instead, at strong coupling the DMFT Green's function turns insulating, suppressing contributions of the residual vertex~\cite{Krien19,Krien19-2}.

We will now analyze and explain in detail the behavior of the different approximations, where we rely on the useful insights of Ref.~\cite{delRe21} into the role of different fluctuations for DMFT's N\'eel temperature. We begin by comparing the SBE approximation (red curve) to the $w$ approximation (black curve), whose difference is particularly transparent on a formal level: The SBE approximation on the top of Fig.~\ref{fig:jib} retains the Hedin vertices, but they are set to $\pm1$ in the $w$ approximation. Hence, remarkably, the Hedin vertices make all the difference between the red and the black curves in Fig.~\ref{fig:phase_diag}. This is even more surprising considering the fact that the SBE approximation is composed of the {\sl same} basic fluctuations as the $w$ approximation ($w^\ch, w^\sz, w^\sing$), and therefore the Hedin vertices can only emphasize or suppress contributions of fluctuations that are already included in the $w$ approximation, nevertheless leading to the substantial differences between the two approximations.

As seems natural, we find that the key difference between the N\'eel temperatures of the $w$ and SBE approximations originates in their different emphasis of the spin fluctuations represented by $w^\sz$. It is shown in Ref.~\cite{delRe21} that the contribution of spin fluctuations to the impurity vertex $f$ works to {\sl enhance} the N\'eel temperature of DMFT. Therefore, in order to get from the black curve in Fig.~\ref{fig:phase_diag} to the red curve, the fermion-spin-boson coupling $\lambda^\sz(\nu,\omega=0)$ needs to slightly suppress the contribution of spin fluctuations at weak coupling~\footnote{The case of weak coupling is complicated by the interplay of different fluctuations with the incoherence introduced by the DMFT self-energy~\cite{delRe21}. One should note that the authors of Ref.~\cite{delRe21} consider a different Bethe-Salpeter equation using the particle-hole-irreducible vertex of the AIM, whereas Eq.~\eqref{eq:BSE} is formulated in terms of the full vertex $f$. This may slightly change the effect of different truncations at the two-particle level.} and hugely enhance it in the strong coupling limit. This quantity is shown in the insets of Fig.~\ref{fig:phase_diag} for three pairs $(U,T)$. Indeed, at weak coupling $\lambda^\sz(\nu,\omega=0)$ is suppressed due to the Kanamori screening~\cite{Krien20}, while at strong coupling it is enhanced by multiple times of its noninteracting value $1$; in both cases these features are localized at small frequencies. In fact, in the deeply insulating regime $\lambda^\sz$ grows without limit, for example, for $U/t=11, T/t\approx0.2$ values as large as $25$ were observed~\footnote{It therefore appears that in a local moment regime $\lambda^\sz(\nu,\omega=0)$ may diverge for small $\nu$ as $T\rightarrow0$. However, the spin polarization $\pi^\sz(\omega\!=\!0)\!=\!\sum_{\nu}g_{\nu}g_{\nu}\lambda^\sz(\nu,\omega\!=\!0)\rightarrow-\frac{1}{U}$  and the effective exchange $J$ in Eq.~\eqref{eq:jeff} remain finite. This requires that the divergence of $\lambda^\sz$ is cancelled by a zero of $g_\nu$, which is plausible because the ground state of the AIM is insulating in the local moment regime. This corrects a previous statement made in the conclusions of Ref.~\cite{Krien19}.}.

Finally, this comparison also explains the behavior of the asymptotic approximation for strong coupling, which lies in the middle between the two other approximations (compare black, purple, and red curves in Fig.~\ref{fig:phase_diag}). This is the case because compared to the $w$ approximation only one of the Hedin vertices $\lambda^\sz(\nu,\omega=0)$ is set to $1$ [cf. Eq.~\eqref{eq:asym}] and thus the effect of this truncation is alleviated. However, as Fig.~\ref{fig:phase_diag} and Eq.~\eqref{eq:jeff} underline, the effective exchange $J$ is mediated by {\sl both} Hedin vertices~\cite{Krien19} and thus also the asymptotic approximation eventually fails at strong coupling. 

\begin{figure}[b]
\begin{center}
\begin{tikzpicture}
\node[anchor=south west,inner sep=0] (image1) at (0,0)
{\includegraphics[width=0.25\textwidth]{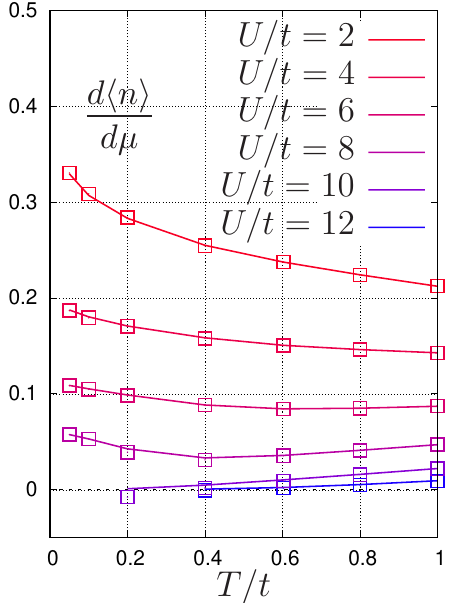}};
\node[anchor=south west,inner sep=0] (image1) at (4.3,0)
{\includegraphics[width=0.25\textwidth]{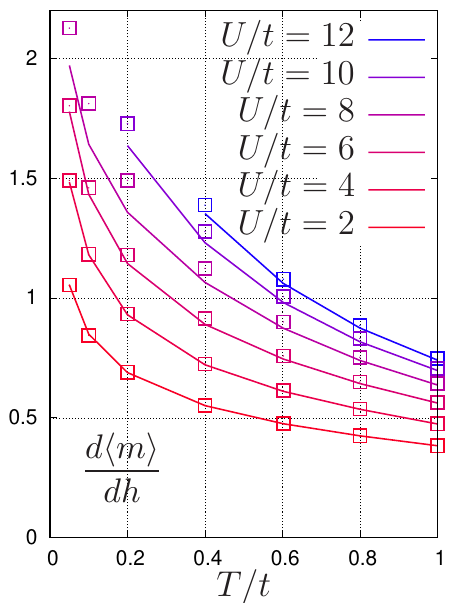}};
\end{tikzpicture}
\end{center}
\caption{Static homogeneous charge (left) and spin (right) susceptibility. Lines show the DMFT result, symbols denote the SBE approximation.}
\label{fig:static_susc_T}
\end{figure}

\section*{DMFT -- Susceptibilities}
As a further benchmark we calculate the DMFT susceptibility using the vertex $F$~\footnote{
In practice we use the formula presented in Ref.~\cite{Krien19} to evaluate the susceptibility $X$, which converges faster with the Matsubara cutoff than Eq.~\eqref{eq:susc}. Using Eq.~\eqref{eq:susc} in the text allows us to discuss DMFT and the LDFA self-energy~\eqref{eq:SDE} on equal footing.} obtained from the Bethe-Salpeter Eq.~\eqref{eq:BSE},
\begin{align}
X^{\alpha}(q)=\sum_{\nu}X^{0}_{\nu}(q)+\sum_{\nu\nu'}X^{0}_{\nu}(q)F^{\alpha}_{\nu\nu'}(q)X^{0}_{\nu'}(q).~\label{eq:susc}
\end{align}
Here, $X^{0}_{\nu}(q)=\sum_{\textbf{k}}G_{k}G_{k+q}$ denotes a bubble of DMFT Green's functions.

We evaluate the static homogeneous charge susceptibility $X^{\ch}(\qv=\mathbf{0},\omega=0)=-\frac{dn}{d\mu}$ and spin susceptibility $X^{\sz}(\qv=\mathbf{0},\omega=0)=-\frac{dm}{dh}$ and compare the SBE approximation to the full DMFT solution (for simplicity we do not consider here the $w$ and asymptotic approximations). The result is shown in Fig.~\ref{fig:static_susc_T}, showing excellent agreement over a wide range of temperatures and interactions, regardless of whether the DMFT solution describes a good/bad metal or a Mott insulating phase. The largest absolute deviations occur in the spin channel for intermediate $U/t\approx8$. Actually, the relative deviations in the charge channel are similar, but the charge susceptibility is vanishingly small in this case.

\begin{figure}
\begin{center}
\begin{tikzpicture}
\node[anchor=south west,inner sep=0] (image1) at (0,0)
{\includegraphics[scale=0.7]{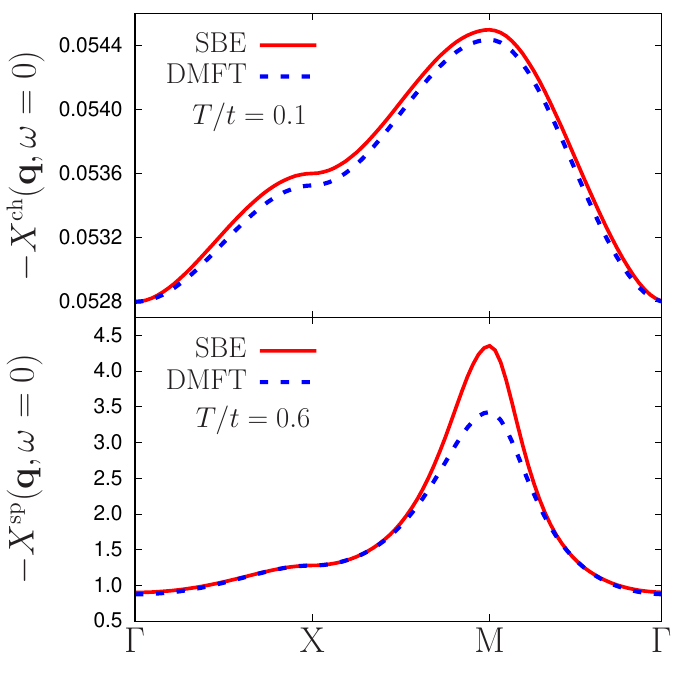}};
\end{tikzpicture}
\end{center}
\caption{Static charge (top) and spin (bottom) susceptibility for $U/t=8$ as a function of momentum \textbf{q}.}
\label{fig:susc_gmx}
\end{figure}

Using the SBE approximation, we also confirm qualitative agreement of the momentum dependence with the DMFT solution in the most delicate coupling regime $U/t=8$. Fig.~\ref{fig:susc_gmx} shows $-X^{\ch/\sz}(\qv,\omega=0)$ for $\qv$ on the high-symmetry path. $-X^\ch$ is shown at a low temperature, where the DMFT solution corresponds to a strongly correlated Fermi liquid (cf. also left panel of Fig.~\ref{fig:static_susc_T}). The spin susceptibility was computed slightly above the N\'eel temperature of the SBE approximation, showing a sizable quantitative difference compared to DMFT but qualitatively correct momentum dependence. Of course, since the SBE approximation has a different N\'eel temperature compared to DMFT, the quantitative difference can be arbitrarily large near this point.

\begin{figure}
\begin{center}
\begin{tikzpicture}
\node[anchor=south west,inner sep=0] (image1) at (0,4.6)
{\includegraphics[width=.5\textwidth]{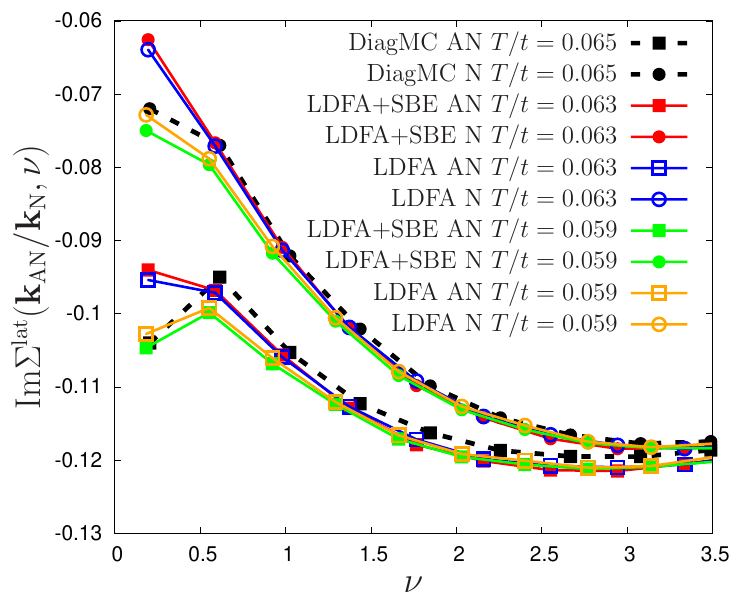}};
\node[anchor=south west,inner sep=0] (image1) at (0.69,0)
{\includegraphics[width=.23\textwidth]{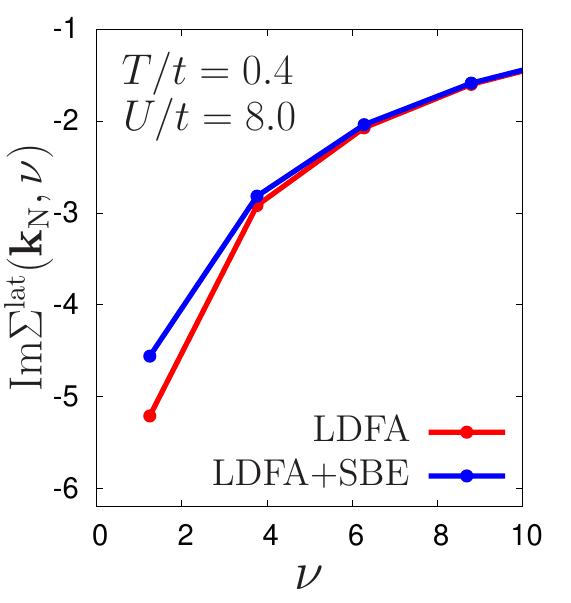}};
\node[anchor=south west,inner sep=0] (image1) at (4.52,0)
{\includegraphics[width=.23\textwidth]{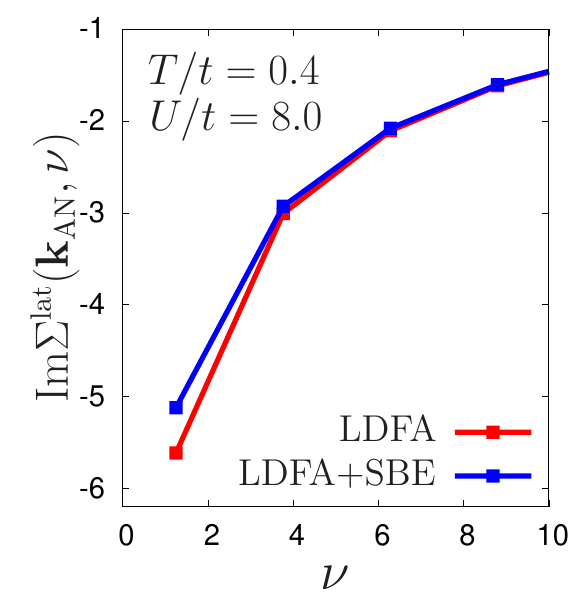}};
\end{tikzpicture}
\end{center}
\caption{Top: Imaginary part of the LDFA self-energy obtained for $U/t=2$ with the temperature above and below $T_{PG}$ (see text). Circles (squares) show the nodal (antinodal) point. Open symbols correspond to LDFA, and solid symbols show LDFA using the SBE approximation for the impurity vertex. Black dashed lines show the DiagMC result of Ref.~\cite{Schaefer21} for $T/t=0.065$. Bottom: LDFA and LDFA+SBE self-energy at intermediate coupling.}
\label{fig:Sigma_dmft_bath}
\end{figure}

\section*{LDFA -- Pseudogap at weak coupling}
As our last benchmark, we use the SBE approximation to evaluate the momentum-dependent self-energy within the ladder dual fermion approach (LDFA). We calculate the dual self-energy $\tilde{\Sigma}$ via the Schwinger-Dyson equation in the following form~\cite{Otsuki14},
\begin{align}
\tilde{\Sigma}(k)=&-\sum_{k'} f^{ch}_{\nu\nu',\omega=0}\tilde{G}(k')\\ \nonumber
&+\frac{1}{4}\sum_{q}\tilde{G}(k+q)[V^{ch}_{\nu\nu'}(q)+3V^{sp}_{\nu\nu'}(q)],\label{eq:SDE}
\end{align}
where $V^{\alpha}$ is defined as
\begin{align}
V^{\alpha}_{\nu\nu'}(q)=\sum_{\nu''} f^{\alpha}_{\nu\nu''\omega}\tilde{X}^{0}_{\nu''}(q)[2F^{\alpha}_{\nu''\nu'}(q)-f^{\alpha}_{\nu''\nu'\omega}].
\end{align}
Note that here $\tilde{G}(k)$ denotes the dual Green's functions dressed with the dual self-energy $\tilde{\Sigma}$. At the end of the calculation the lattice self-energy is obtained using the relation $\Sigma^{\text{latt}}_k=\Sigma_\nu+\tilde{\Sigma}_k(1+g_\nu\tilde{\Sigma}_k)^{-1}$, where $\Sigma_\nu$ is the local self-energy of the AIM~\footnote{The AIM corresponds to the DMFT solution, that is, we keep the hybridization function $\Delta$ in Eq.~\eqref{eq:aim} fixed to its DMFT value.}. As we are interested in the low-energy behavior of the self-energy we use a physical cutoff $\nu^\text{max}=15t$ for improved performance.

We apply the LDFA to the Hubbard model~\eqref{eq:hubbard} at weak coupling $U/t=2$, this regime was recently investigated in Ref.~\cite{Schaefer21} using diagrammatic Monte Carlo (DiagMC,~\cite{Prokofiev98}). Top panel of Fig.~\ref{fig:Sigma_dmft_bath} shows the LDFA result for $\Sigma^{\text{latt}}(\kv_N/\kv_{AN},\nu)$ (open symbols), where $\kv_N=(\frac{\pi}{2},\frac{\pi}{2})$ and $\kv_{AN}=({\pi},0)$ correspond to the nodal and antinodal point, respectively. The LDFA captures the opening of the pseudogap at a temperature $T_{PG}\approx0.059t$, which lies below the numerically exact value of $0.065t$ (see Ref.~\cite{Schaefer21} and black dashed lines in Fig.~\ref{fig:Sigma_dmft_bath}). To achieve a better quantitative agreement with DiagMC parquet diagrams need to be taken into account~\cite{Krien20-3}. The top panel 
also shows the LDFA result when the SBE approximation is used for the impurity vertex $f$ (closed symbols), in quantitative agreement with the original LDFA. Hence, the SBE approximation recovers signature physics also of diagrammatic extensions of DMFT~\cite{Rohringer18}.

\section*{LDFA -- Intermediate coupling}
We benchmark the combination of LDFA and SBE approximation also for $U/t=8$ and $T/t=0.4$. The bottom panels of Fig.~\ref{fig:Sigma_dmft_bath} show a reasonable agreement of the self-energy at node and antinode, which are both insulating in this regime. Interestingly, we observe that the leading eigenvalue of the matrix~\eqref{eq:matrix_ev} is larger for LDFA+SBE ($\lambda^\text{max}\approx0.81$) compared to LDFA ($\lambda^\text{max}\approx0.75$), while the absolute value of the self-energy is smaller. This implies a nontrivial relationship between the frequency structure of the impurity vertex and the feedback on the self-energy. This is not obvious from comparison of LDFA self-energies at different temperatures, where it may appear that the feedback is mainly controlled by the leading eigenvalue.

However, let us note that the essential physics of the Hubbard model in this interaction regime corresponds to localized spins interacting via the effective exchange. Hence, it is plausible that the mere self-consistent feedback of spin fluctuations on the electronic self-energy captured by the ladder approximation is {\sl not} the salient physical effect, but that instead self-renormalization of (i.e., interaction between) spin fluctuations should be taken into account, see, e.g., Ref.~\cite{Katanin21}. This requires more sophisticated diagrammatic resummation schemes, such as the parquet diagrams~\cite{Bickers92,Krien20-3,Krien21}.

\begin{figure}
\includegraphics[scale=0.7]{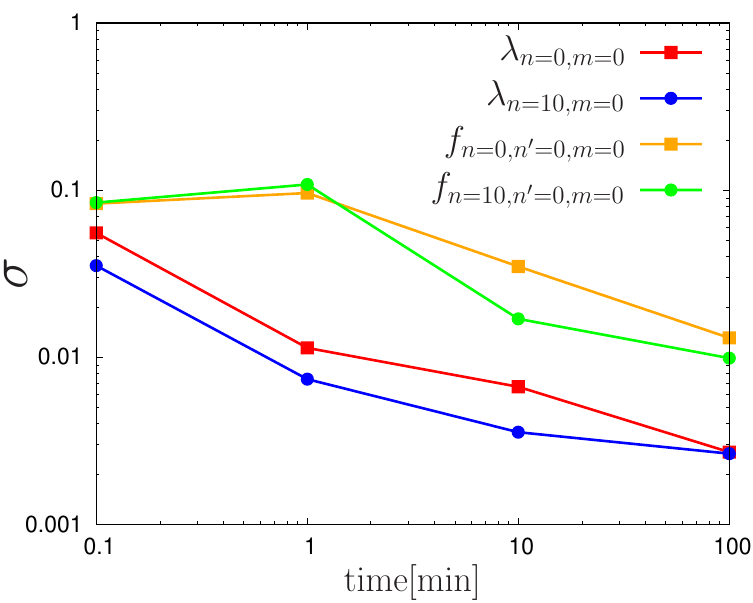}
\caption{Relative standard deviation of indicated matrix elements of three- and four-leg vertices $\lambda$ and $f$, respectively, as functions of the measurement time using 192 CPUs.}
\label{fig:deviation}
\end{figure}

\section*{Computational cost}
The SBE approximation depicted in Fig.~\ref{fig:jib} avoids the measurement of the four-point vertex function $f(\nu,\nu',\omega)$, which is the computationally most demanding task in calculations of the DMFT susceptibility and in ladder extensions of DMFT~\cite{Galler19}. The measurement of the three-leg vertices $\lambda(\nu,\omega)$ is cheaper, however, the reduction of computational cost depends on details of the implementation, on the autocorrelation time that is specific both to the observable and to the physical regime, and on the method of measurement (e.g. segment or worm sampling). Fig.~\ref{fig:deviation} provides an illustrative example of a DMFT calculation at $U/t=8, T/t=0.1$ where $f^{\ch/\sz}$ and $\lambda^{\ch/\sz}$ were measured on $142$ fermionic and $142$ bosonic frequencies using the segment solver with improved estimators presented in Refs.~\cite{ALPS2,Hafermann12}. We used 192 CPUs for the measurements. Fig.~\ref{fig:deviation} shows the relative standard deviations of $f(\nu_0,\nu_0,\omega_0), f(\nu_{10},\nu_0,\omega_0), \lambda(\nu_0,\omega_0)$, and $\lambda(\nu_{10},\omega_0)$ against the total measurement time in minutes. Apparently, in this setup, obtaining $f$ and $\lambda$ to similar accuracy requires roughly a one hundred times longer measurement time for $f$.

Further, the SBE approximation also preserves memory and disk space. The first line of Table~\ref{tab:diskspace} shows the size of $\lambda$ and $f$ calculated above, corresponding to the single-band Hubbard model~\eqref{eq:hubbard}. The second and third line indicate corresponding sizes for a three-orbital Hubbard-Kanamori Hamiltonian either with or without SU(2) symmetry~\cite{Wallerberger19,Galler19}. Notice that the center column corresponds to the {\sl particle-hole} quantities $\lambda^{\ch/\sz}$ [cf. Appendix Eq.~\eqref{eq:3-vertex_ph}]. The singlet vertex $\lambda^\sing$ [cf. Eq.~\eqref{eq:3-vertex_pp}] is of similar size and, furthermore, the multi-orbital case requires in general also the measurement of a triplet component (it vanishes in the single-band case due to the Pauli principle), which needs to be taken into account in the derivation of the SBE decomposition~\cite{Krien19-4}.

\begin{table}
\centering
\begin{tabular}{ |c|c|c| }
 \hline
  & ~$\lambda$~ & ~$f$~ \\ 
 \hline
 1 orb. + SU(2) & 2.13 MB & 92.7 MB \\ 
 3 orb. + SU(2) & 44.73 MB & 1946.7 MB \\
 3 orb. w/o SU(2) & 134.19 MB & 5840.1 MB \\
 \hline
\end{tabular}
\caption{\label{table} Size of three- and four-leg vertices $\lambda$ and $f$ for 142 fermionic/bosonic frequencies. First line shows the single-band case with SU$(2)$ symmetry, second and third line correspond to a three orbital Hubbard-Kanamori model}
\label{tab:diskspace}
\end{table}

\section*{Properties of the residual vertex}
The benchmarks in the previous sections confirm that the SBE approximation recovers the two-particle physics of DMFT and its diagrammatic extensions from weak to strong coupling quantitatively. 
While 
at strong coupling an analytical argument 
favors the SBE approximation over the $w$ and asymptotic approximations,
one may still doubt whether it is preferable in general.

In this section we provide numerical evidence that, for the purpose of a parametrization of the vertex, the organizing principle of the SBE approximation should be favored over that of the asymptotic approximation: At all energy scales, the local vertex corrections should be parametrized through bosonic fluctuations and their coupling to the fermions, rather than through a combination of asymptotic expressions~\footnote{It may seem that the guiding principle of the asymptotic approximation is also to parametrize the vertex in terms of bosons and their coupling $\lambda$ to fermions. However, this would not be an accurate description, because one vertex $\lambda$ is always set to its asymptotic value $\pm1$, cf. Eq.~\eqref{eq:asym}, which at small frequencies does not correspond to the fermion-boson coupling of the interacting system.}. To show this, we compute the residual vertices (rest functions) of, respectively, the SBE decomposition and of the vertex asymptotics and compare them at small frequencies.

The residual vertex $\varphi^\text{Uirr}$ of the SBE decomposition is defined as the set of vertex diagrams which do not contain insertions of the bare interaction $U$~\cite{Krien19-4}. On the other hand, we refer to the residual vertex of the vertex asymptotics as $\varphi^{\text{as}}$. To get at a clearer picture what $\varphi^\text{as}$ is, we pinpoint its difference to $\varphi^\text{Uirr}$ as follows. First, we recall that the asymptotic approximation can be obtained from the SBE approximation by combining different high-frequency limits of the SBE diagrams as in Eq.~\eqref{eq:asym}. Let us determine explicitly what is neglected in the asymptotic approximation, for example, looking at $\nabla^{ph}$ defined in Eq.~\eqref{eq:SBEvertices_ph} [$\alpha=\ch,\sz$],
\begin{align}
\nabla^{ph,\alpha}_{\nu\nu'\omega}=&{\lambda}^{\alpha}_{\nu\omega} w_{\omega}^{\alpha}\lambda_{\nu'\omega}^{\alpha}\notag\\
=&{\lambda}^{\alpha}_{\nu\omega} w_{\omega}^{\alpha}
+w_{\omega}^{\alpha}\lambda_{\nu'\omega}^{\alpha}
-w_{\omega}^{\alpha}\notag\\
&+({\lambda}^{\alpha}_{\nu\omega}-1)w_{\omega}^{\alpha}(\lambda_{\nu'\omega}^{\alpha}-1).\notag
\end{align}
In the asymptotic approximation the term in the last line is neglected (that is, absorbed into $\varphi^\text{as}$). Similar terms arise from taking the asymptotic limits of $\nabla^{\overline{ph}}$ and of $\nabla^{{pp}}$. As a result, the difference between the residual vertices $\varphi^\text{Uirr}$ and $\varphi^\text{as}$ is that the latter contains remainders of the form $R=(\lambda\mp1)w(\lambda\mp1)$. They correspond to the boson $w$ coupled to fermions only via the nontrivial (interacting) part of the fermion-boson coupling $\lambda$. In the following we evaluate $\varphi^\text{Uirr}$ and $\varphi^\text{as}$ in a nontrivial parameter regime and pay special attention to the behavior of the remainders $R$.

\begin{figure}
\begin{center}
\begin{tikzpicture}
\node[anchor=south west,inner sep=0] (image1) at (0,0)
{\includegraphics[width=0.25\textwidth]{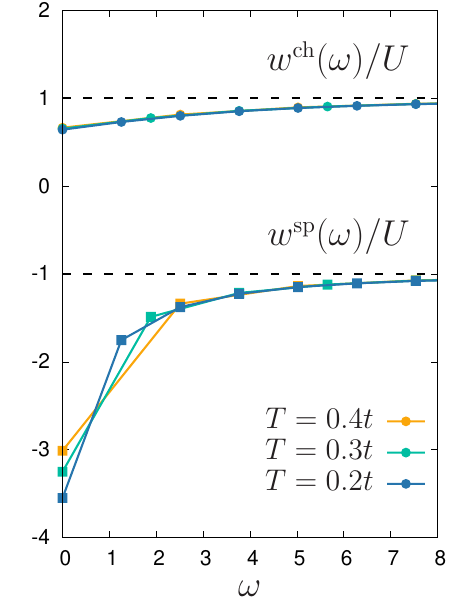}};
\node[anchor=south west,inner sep=0] (image1) at (4.3,0)
{\includegraphics[width=0.25\textwidth]{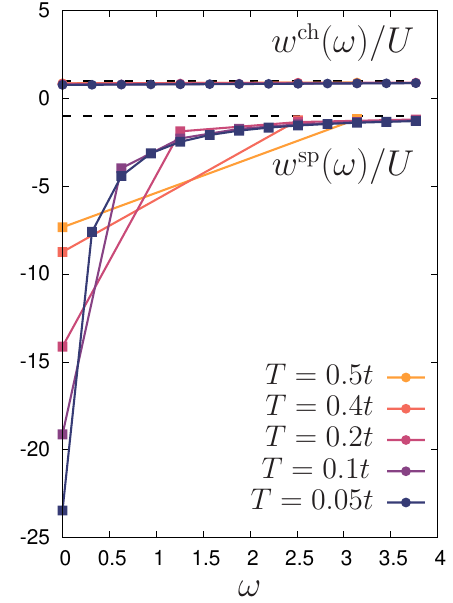}};
\node[anchor=south west,inner sep=0] (image1) at (0,6.2)
{\includegraphics[width=0.25\textwidth]{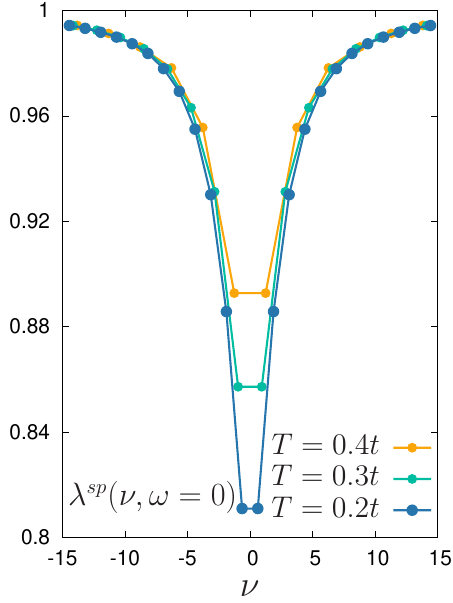}};
\node[anchor=south west,inner sep=0] (image1) at (4.3,6.2)
{\includegraphics[width=0.25\textwidth]{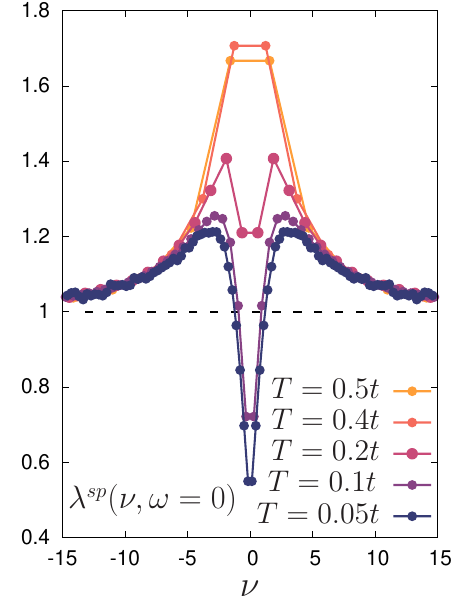}};
\draw[->,thick] (6.1,8.55) -- (6.5,8.95);
\end{tikzpicture}
\end{center}
\caption{Top: Local fermion-spin-boson coupling ($\omega=0$) corresponding to DMFT calculations at $U/t=4$ (left) and $U/t=8$ (right) for various temperatures. Bottom: Screened interaction for the same parameters.}
\label{fig:vertices}
\end{figure} 

\begin{figure*}
\begin{center}
\begin{tikzpicture}
\node[anchor=south west,inner sep=0] (image1) at (0,0)
{\includegraphics[width=0.95\textwidth]{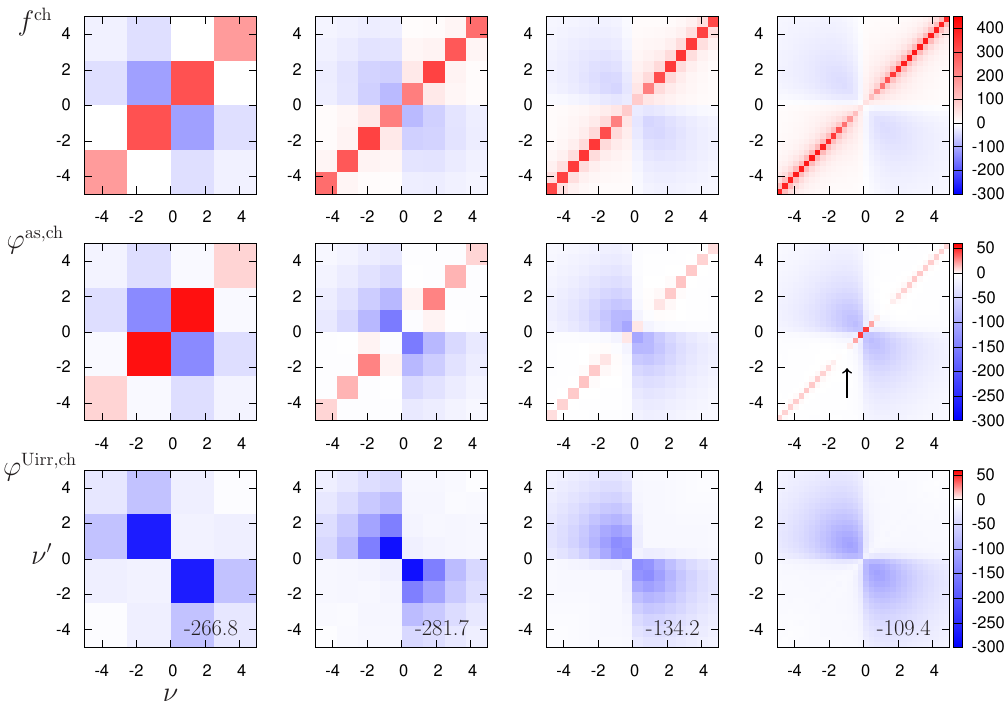}};
\end{tikzpicture}
\end{center}
\vspace{-.5cm}
\caption{\label{fig:vertex_ch}
Top: Full charge vertex $f^\ch(\nu,\nu',\omega=0)$ for $U/t=8$ focused on small $\nu,\nu'$. From left to right panels correspond to $T/t=0.4, 0.2, 0.1$, and $0.05$. Center and bottom: Residual vertices of the vertex asymptotics (center) and of the SBE decomposition (bottom). The arrow indicates an `accidental' cancellation of the diagonal by the vertex asymptotics, which occurs when $\lambda^\sz$ crosses its asymptotic value $1$ (see text and arrow in Fig.~\ref{fig:vertices}). Color schemes are consistent in each row. Numbers in the bottom panels indicate the global minimum of $\varphi^{\text{Uirr},\ch}$.}
\end{figure*}

As the test ground we choose $U/t=8$ and perform DMFT calculations at temperatures $T/t=0.5, 0.4, 0.2, 0.1,$ and $0.05$. This sequence is interesting in DMFT because at high temperature the self-consistent AIM exhibits a local moment, but crosses over into a Fermi liquid at low temperature. This change manifests itself in the impurity correlation functions, as can be observed in the right panels of Fig.~\ref{fig:vertices} which show the Hedin vertex $\lambda^\sz(\nu,\omega=0)$ and the screened interaction $w(\omega)$.

The former shows a strong enhancement in the local moment regime at high temperature, and a similar strong suppression in the Fermi liquid at low temperature. Actually, this is comparable to the sequence of Hedin vertices shown in the insets of Fig.~\ref{fig:phase_diag}, where the interaction was changed to switch between the two regimes. In the present case, however, it is the peculiar temperature dependence of the DMFT approximation that drives the crossover. The latter can also be observed in the static screened interaction $w^\sz(\omega=0)$, which roughly doubles going from $T/t=0.4$ to $T/t=0.2$, consistent with a local moment $\propto\beta$, whereas at low temperature the increase of $w^\sz(\omega=0)$ is markedly slower. On the other hand, $w^\ch$ is tiny in comparison, and hence we can ignore this quantity (and $\lambda^\ch$) in the following~\footnote{Due to particle-hole symmetry the singlet quantities $w^\sing$ and $\lambda^\sing$ can be obtained from the charge channel~\cite{Krien19-4} and hence they are also small and can be ignored.}.

To fully characterize the setting, let us also compare to a case of weaker coupling $U/t=4$ shown in the left panels of Fig.~\ref{fig:vertices}. In this case $\lambda^\sz(\nu,\omega=0)$ does not show an enhancement at high temperature. Interestingly, the comparison to the right panels suggests that the screening of fermions from spin fluctuations, mediated by a suppression of $\lambda^\sz(\nu,\omega=0)$, is not {\sl per se} a feature of the weak coupling regime, but rather of the Fermi liquid phase (although in the strongly correlated Fermi liquid the mechanism may be more multifaceted than the simple Kanamori screening at weak coupling~\cite{Krien20}).

At the same time, however, we note the qualitative difference between the Fermi liquid regimes at $U/t=4$ and $U/t=8$: In the latter case $\lambda^\sz(\nu,\omega=0)$ is strongly suppressed at very small frequencies (see e.g. $T/t=0.05$), but it shows an enhancement for intermediate frequencies which is not there for $U/t=4$. This indicates that for $U/t=8$ only low-energetic fermions are screened from spin fluctuations and reminds of the `onion shape' of the generalized charge susceptibility recently observed in Ref.~\cite{Chalupa21}.

\begin{figure*}
\begin{center}
\begin{tikzpicture}
\node[anchor=south west,inner sep=0] (image1) at (0,0)
{\includegraphics[width=0.95\textwidth]{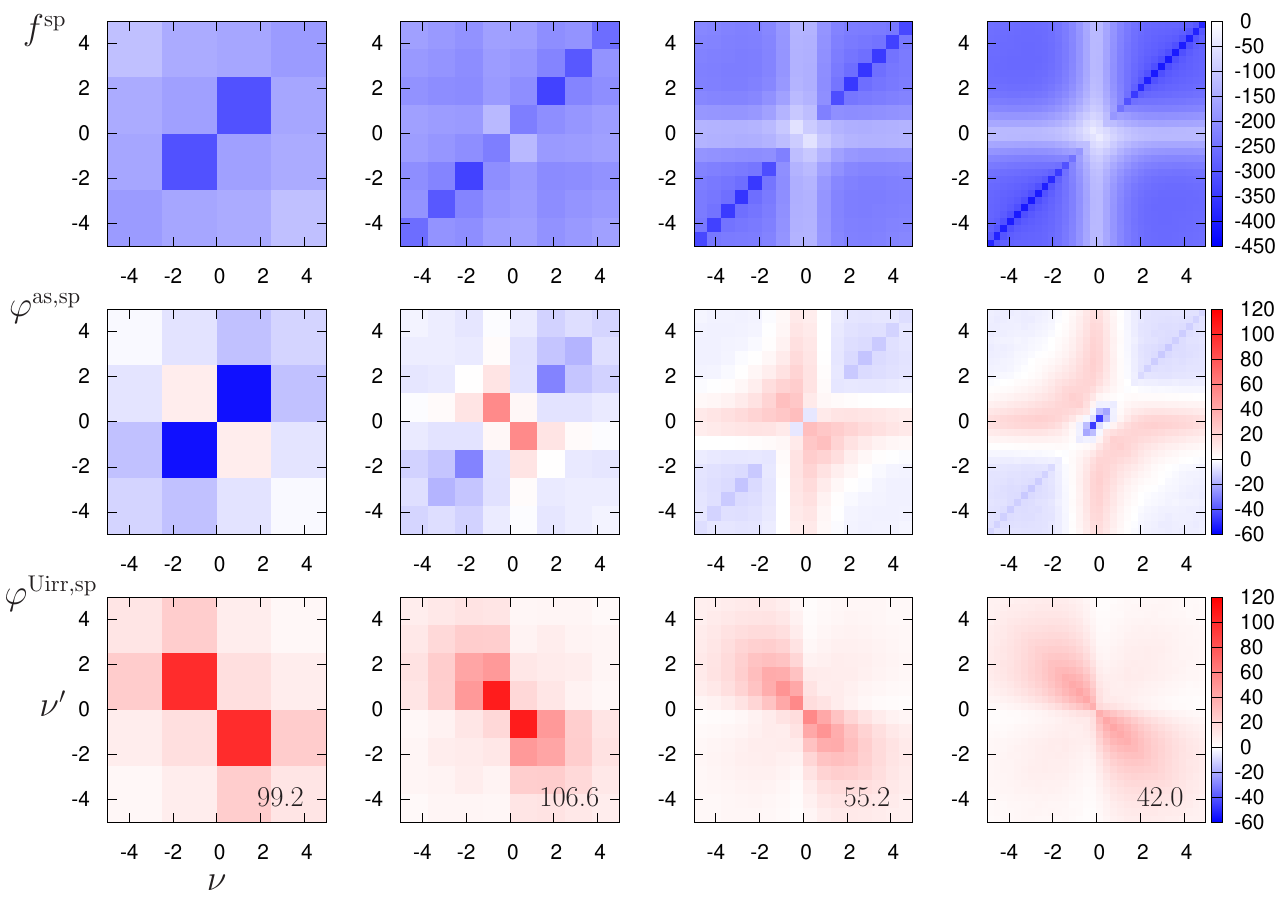}};
\end{tikzpicture}
\end{center}
\vspace{-.5cm}
\caption{\label{fig:vertex_sp}
Full (top) and residual (center, bottom) vertices of the spin channel for the same parameters as in Fig.~\ref{fig:vertex_sp}. Numbers in the bottom panels indicate the global maximum of $\varphi^{\text{Uirr},\sz}$.}
\end{figure*}

In summary, we have identified a crossover from a local moment to a Fermi liquid phase that manifests itself as a qualitative change in the quantities $\lambda$ and $w$, which are used to formulate the SBE approximation and the asymptotic approximation. Next, we observe the behavior of the residual vertices $\varphi^\text{Uirr}$ and $\varphi^\text{as}$ at the crossover.

The bottom panels of Fig.~\ref{fig:vertex_ch} show the residual charge vertex $\varphi^{\text{Uirr},\ch}(\nu,\nu',\omega=0)$ of the SBE decomposition for different temperatures, where the energy window for $\nu$ and $\nu'$ is consistent. In the local moment regime (Fermi liquid) $\varphi^{\text{Uirr},\ch}$ increases (decreases) in absolute magnitude as $T$ is lowered (see numbers in the bottom panels), but otherwise exhibits a remarkable uniformity, showing the same features, as it were, in different resolutions. Indeed, in Fig.~\ref{fig:vertex_sp} we find the same behavior in $\varphi^{\text{Uirr},\sz}(\nu,\nu',\omega=0)$, albeit with the opposite sign.

As observed previously~\cite{Krien19-4}, $\varphi^{\text{Uirr}}(\omega=0)$ has no significant features in the two sectors where $\text{sgn}(\nu)=\text{sgn}(\nu')$, indicating an almost perfect cancellation of the SBE diagrams $\nabla^\text{SBE}$ with the full vertex $f$ in these sectors. On the other hand, there are features located in the two sectors where $\text{sgn}(\nu)=-\text{sgn}(\nu')$, respectively, provided $\nu$ and $\nu'$ are both small. Consistently, and also at smaller interaction~\cite{Krien19-4}, these features correspond to attraction in the charge channel, $\varphi^{\text{Uirr},\ch}(\omega=0)<0$, and repulsion in the spin channel, $\varphi^{\text{Uirr},\sz}(\omega=0)>0$.

The origin and physical nature of these features are unknown. One may speculate that they are related to particle-particle scatterings, because the secondary diagonal $\nu+\nu'\approx0$ that crosses through the sectors with $\text{sgn}(\nu)=-\text{sgn}(\nu')$ is associated with the singlet pairing channel~\cite{Rohringer12}. However, $\varphi^{\text{Uirr}}$ does not exhibit any resonances for a particular frequency combination. The simple structure of $\varphi^{\text{Uirr}}$, which does not change qualitatively at the crossover from local moment regime to Fermi liquid, suggests that it could represent two-particle correlations which play a similar role in both regimes. A candidate explanation is, hence, that $\varphi^{\text{Uirr}}$ captures, in the strongly correlated Fermi liquid, the {\sl pre-formed} local moment.

The described phenomenology is in strong contrast with the full vertex $f^\ch$ drawn in the top panels of Fig.~\ref{fig:vertex_ch}, which exhibits both repulsive and attractive features, with some cancellations between them for small $\nu$ and/or $\nu'$ that vary with the temperature. In fact, $\varphi^{\text{Uirr},\ch}$ represents in essence the attractive features of $f^\ch$, whereas repulsive single-boson exchange contributes a resonant structure on the diagonal $\nu-\nu'\approx0$. On the other hand, in the spin channel the repulsive contribution of $\varphi^{\text{Uirr},\sz}$ to $f^\sz$ is cancelled out completely by attractive single-boson exchange~\footnote{It is plausible that the N\'eel temperature of the SBE approximation is higher than that of DMFT because the neglected $\varphi^{\text{Uirr},\sz}$ is repulsive, cf. Fig.~\ref{fig:phase_diag}.}, see Fig.~\ref{fig:vertex_sp}. Also here, the residual vertex $\varphi^{\text{Uirr},\sz}$ does not exhibit resonant features.

Next, we examine the residual vertex $\varphi^\text{as}$ of the vertex asymptotics drawn in the second row of the panels in Fig.~\ref{fig:vertex_ch} and~\ref{fig:vertex_sp}, respectively. Since this quantity corresponds to a superset of the diagrams contained in $\varphi^\text{Uirr}$, it is not surprising that the features of $\varphi^\text{Uirr}$ can also be observed in $\varphi^\text{as}$. However, they overlap with remainder structures $R=(\lambda\mp1)w(\lambda\mp1)$, as explained in the beginning of this section. The remainders necessarily decay at high frequencies, where the SBE and asymptotic approximations become equivalent, however, Figs.~\ref{fig:vertex_ch} and~\ref{fig:vertex_sp} focus on small frequencies.

A prominent remainder structure is visible on the diagonal of the residual charge vertex $\varphi^{\text{as},\ch}$ in Fig.~\ref{fig:vertex_ch}, both at high and at low temperatures. We obtain an expression for this feature as follows. The remainder stemming from the {\sl horizontal} particle-hole channel is given as,
\begin{align}
R^{ph,\alpha}_{\nu\nu'\omega}=(\lambda^\alpha_{\nu\omega}-1)w^\alpha_\omega(\lambda^\alpha_{\nu'\omega}-1),
\end{align}
 where $\alpha=\ch,\sz$. Since $w^\ch/U\approx1$ is comparatively tiny in the considered regime (cf. Fig.~\ref{fig:vertices}, bottom right), $R^{ph,\ch}$ is negligible and hence not easily visible in $\varphi^{\text{as},\ch}$. However, a remainder $R^{\overline{ph}}$ contributes to $\varphi^{\text{as}}$ also in the {\sl vertical} particle-hole channel. This remainder arises from the crossing relation~\eqref{eq:SBEvertices_vph} and is {\sl not} small, because the crossing relation mixes the charge and spin flavors. Neglecting a contribution from $R^{ph,\ch}$ we obtain,
\begin{align}
R^{\overline{ph},\ch}_{\nu,\nu',\omega=0}\propto(\lambda^\sz_{\nu,\nu'-\nu}-1)w^\sz_{\nu'-\nu}(\lambda^\sz_{\nu,\nu'-\nu}-1).
\end{align}
As expected, this quantity decays for large $\nu$ or $\nu'$ where the Hedin vertex approaches $1$, however, for small $\nu\approx\nu'$ it contributes to $\varphi^{\text{as},\ch}$, that is, the red features in the second row of panels in Fig.~\ref{fig:vertex_ch}.

Comparison with Fig.~\ref{fig:vertices} elucidates the peculiar structure of this feature: Firstly, the screened interaction $w^\sz(\omega)$ decays quickly for finite $\omega$, hence $R^{\overline{ph},\ch}$ displays a strong resonance only on the diagonal of $\varphi^{\text{as},\ch}$. Secondly, at low temperature (e.g. $T=0.05$) the Hedin vertex $\lambda^\sz_{\nu,\omega=0}$ changes for $\nu\approx\pm1$ from suppression ($\lambda^\sz<1$) to enhancement ($\lambda^\sz>1$). Therefore, $R^{\overline{ph},\ch}$ vanishes at this point. Indeed, the red feature in $\varphi^{\text{as},\ch}$ disappears near $\nu\approx\pm1$ (see arrows in Figs.~\ref{fig:vertices} and~\ref{fig:vertex_ch}), but reemerges again at larger frequencies.

The described effect serves as strong evidence that in the asymptotic approximation cancellations with the full vertex $f$ occur at small frequencies only {\sl accidentally}. In contrast, we have confirmed in a nontrivial setting, featuring a crossover from local moment to Fermi liquid phase, that the SBE diagrams capture the resonant features of $f$ even at the smallest frequencies. Therefore, we conclude that the formal construction of the SBE decomposition retains a {\sl physical} correspondence to the full vertex function at all energy scales.

\section*{Conclusions}\label{sec:conclusion}
We benchmarked an efficient approximation scheme for the local vertex function of the Anderson impurity model in terms of single-boson exchange (SBE,~\cite{Krien19-4}). The SBE approximation is based on the proven and successful organizing principle to parametrize two-particle correlations in terms of bosonic fluctuations and their coupling to fermions~\cite{Denz20,Husemann09,Husemann12,Eberlein_thesis,Krien20,Vandelli20,Harkov21}. We presented analytical arguments for the validity of the SBE approximation at weak and strong coupling and emphasized the importance of the fermion-boson coupling (Hedin vertex) in both limits. This quantity captures the Kanamori screening at weak coupling and strongly enhances the contribution of spin fluctuations to the vertex function at strong coupling. Hence, in the latter case, setting this quantity to its noninteracting value leads to qualitatively wrong results. This may be of relevance for the extension of recent diagrammatic studies of the optical conductivity~\cite{Worm20,Werner21} to the strong coupling regime.

In our numerical applications to the single-band Hubbard model the SBE approximation recovered the two-particle physics of DMFT and of the ladder dual fermion approach to good quantitative accuracy. As a result, the computational cost of evaluating correlation functions of the auxiliary Anderson impurity model was reduced to a level comparable with the TRILEX~\cite{Ayral15} and dual TRILEX approaches~\cite{Stepanov19-2,Harkov21}, however, the Bethe-Salpeter equation still needs to be solved. In the estimation of the N\'eel temperature of DMFT the SBE approximation does not reach the excellent accuracy and efficiency of the strong-coupling limit (SCL) formula for the static spin susceptibility of DMFT~\cite{Otsuki19}, but it has a wider range of applicability, for example, to charge excitations and two-particle correlations in general. These results call for the generalization of the SBE approximation to multi-orbital settings, where it may be applied in cases where the computational bottleneck is the evaluation of the vertex function of the impurity model~\cite{Galler19,Kaufmann21,vanLoon21}. The SBE approximation may also be combined with the efficient formula for the DMFT polarization~\cite{Krien19} and other schemes~\cite{Wallerberger20} that help to reduce the computational cost of solving the Bethe-Salpeter equation.

Our results are also of relevance to the development of unbiased methods such as the fRG~\cite{Metzner12,Dupuis21} and parquet schemes~\cite{Bickers04}. We compared the SBE approximation to a widely used parametrization of the vertex based on its asymptotic limits~\cite{Wentzell20}, which are a crucial ingredient of recent progress in the theory of vertex corrections and two-particle excitations, see, for example, Refs.~\cite{Li16,Krien19,Krien19-4,Krien20,Krien20-2,Hille20,Bonetti20}. We showed in a nontrivial crossover regime from a local moment phase to the Fermi liquid that the SBE decomposition generalizes the vertex asymptotics in a physically meaningful way to small frequencies. In particular, it retains the correspondence to all resonant features of the full vertex function down to the smallest energy scales. As is natural for a high-frequency limit, the approximation based on vertex asymptotics does not show this correspondence. Therefore, resonant features remain in the rest function of the vertex asymptotics, which in the case of long-ranged spin-density wave correlations may carry a strong momentum dependence. The rest function of the SBE decomposition is short-ranged even in this case~\cite{Krien20-2,Krien20-3} and, as a result, its momentum dependence can be captured with only a small number of form factors~\cite{Eckhardt20,Krien20}. It seems therefore promising to include the SBE decomposition into state-of-the-art fRG~\cite{Bonetti21} and parquet schemes~\cite{Krien20,Krien20-2}.

Finally, we noted that, in the considered cases, the residual vertex of the SBE decomposition has a remarkably simple frequency structure, which does not change qualitatively at the crossover from the local moment regime to the Fermi liquid, or by changing the interaction. Based on this, we speculate that this vertex represents a connecting element of the two regimes, namely, the pre-formed local moment, either screened or unscreened, respectively. Indeed, the full two-particle information can be reconstructed from the residual vertex~\cite{Krien19-3}, including the corresponding fingerprints of the local moment~\cite{Chalupa21}. However, further investigation is required, for example, an intriguing option is to study the residual vertex of the SBE decomposition on the real axis~\cite{Kugler21,Lee21}.

\acknowledgments
We thank F. \v{S}imkovic for providing the DiagMC data and S. Andergassen, P. Bonetti, K. Held, C. Hille, F. Kugler, J. Mravlje, L. Del Re, G. Rohringer, A. Toschi, D. Vilardi, M. Wallerberger for fruitful discussions and comments.  We acknowledge support from the European Research Council through the Synergy Grant No. 854843 - FASTCORR (A.I.L., V.H.), from the Austrian Science Fund (FWF) through Projects No. P32044 and No. P30997 (F.K.) and from the North-German Supercomputing Alliance (HLRN) under the Project No.~hhp00042 (A.I.L, V.H.).

\appendix

\section{Impurity correlation functions}\label{app:defs}
In this work we employ two different methods (DMFT and LDFA) which both use the vertex function of the AIM, and the aim of this work is to test efficient parametrizations of this quantity.
However, for comparison we also obtain the {\sl exact} vertex function from the four-point correlation function, which is defined as,
\begin{align}
g^{(4),\alpha}_{\nu\nu'\omega}=-\frac{1}{2}\sum_{\sigma_i}s^{\alpha}_{\sigma'_{1}\sigma^{}_{1}}s^{\alpha}_{\sigma'_{2}\sigma^{}_{2}}\langle c^{}_{\nu\sigma^{}_{1}}c^{*}_{\nu+\omega,\sigma'_{1}}c^{}_{\nu'+\omega,\sigma^{}_{2}}c^{*}_{\nu'\sigma'_{2}}\rangle,
\end{align}
where $s^{\alpha}$ are the Pauli matrices and the label $\alpha=\ch,\sz$ denotes the charge and spin channel, respectively. The vertex function is obtained by subtracting the disconnected parts and removing four Green's function legs from the four-point correlation function $g^{(4)}$, 
\begin{align}
f^{\alpha}_{\nu\nu'\omega}=\frac{g^{(4),\alpha}_{\nu\nu'\omega}-\beta g_{\nu}g_{\nu+\omega}\delta_{\nu\nu'}+2\beta g_{\nu}g_{\nu'}\delta_{\omega}\delta_{\alpha,ch}}{g_{\nu}g_{\nu+\omega}g_{\nu'}g_{\nu'+\omega}}.
\label{eq:4-vertex}
\end{align}
The three-point correlation function is defined as,
\begin{align}
g^{(3),\alpha}_{\nu\omega}=-\frac{1}{2}\sum_{\sigma\sigma'}s^{\alpha}_{\sigma'\sigma^{}}\langle c^{}_{\nu\sigma^{}}c^{*}_{\nu+\omega,\sigma'}\rho^{\alpha}_{\omega}\rangle,
\end{align}
where $\rho^{\ch}=n_{\uparrow}+n_{\downarrow}$ and $\rho^{\sz}=n_{\uparrow}-n_{\downarrow}$ are the charge and spin densities. The Hedin vertex is obtained from $g^{(3)}$ as,
\begin{align}
{\lambda}^{\alpha}_{\nu\omega}=\frac{g^{(3),\alpha}_{\nu\omega}+\beta g_{\nu}\av{n}\delta_{\omega}\delta_{\alpha,\ch}}{g_{\nu}g_{\nu+\omega}w^{\alpha}_{\omega}/U^{\alpha}},
\label{eq:3-vertex_ph}
\end{align}
where $w^{\alpha}_{\omega}$ and $U^\alpha$ are defined below and
\begin{align}
{\lambda}^{\sing}_{\nu{\omega}}=\frac{\av{c_{\nu\uparrow}c_{{\omega}-\nu,\downarrow}\rho^{+}_{{\omega}}}}{g_{\nu}g_{{\omega}-\nu}w^{\sing}_{{\omega}}/U^{\sing}}.
\label{eq:3-vertex_pp}
\end{align}

The charge, spin, and singlet impurity susceptibilities are defined as,
\begin{align}
\chi^{\alpha}_{\omega}=&-\av{\rho^{\alpha}_{-\omega}\rho^{\alpha}_{\omega}}+\beta\av{n}\av{n}\delta_{\omega}\delta_{\alpha,\ch},\\
\chi^{\sing}_{{\omega}}=&-\av{\rho^{+}_{-{\omega}}\rho^{-}_{{\omega}}},
\label{eq:susc_singlet}
\end{align}
and $\rho^{+}=c^{*}_{\uparrow}c^{*}_{\downarrow}$ and $\rho^{-}=c_{\downarrow}c_{\uparrow}$ describe creation/annihilation of electron pairs, respectively. 
The label `$\sing$' in Eq.~(\ref{eq:susc_singlet}) refers to the singlet pairing channel. The screened interaction is obtained from the susceptibility as,
\begin{eqnarray}
    w^\alpha(\omega)=U^\alpha+\frac{1}{2}U^\alpha\chi^\alpha(\omega)U^\alpha,
    \label{eq:w}
\end{eqnarray}
for each corresponding channel $\alpha=\ch,\sz,\sing$. Finally, the bare interaction is defined as,
\begin{align}
    U^\ch=+U,\;\;\;U^\sz=-U,\;\;\;U^\sing=+2U.
    \label{eq:bareint}
\end{align}

\bibliography{main}

\end{document}